\documentclass[a4paper,fleqn,usenatbib]{mnras}

\usepackage{newtxtext,newtxmath}

\usepackage[T1]{fontenc}
\usepackage{ae,aecompl}

\usepackage{graphicx}	
\usepackage{amsmath}	
\usepackage{amssymb}	

\title[Triggering mixing and deceleration in FRI jets]{Triggering mixing and deceleration in FRI jets: a solution}

\author[M. Perucho et al.]{
Manel Perucho,$^{1,2}$,\thanks{E-mail: manel.perucho@uv.es}
\\
$^{1}$Departament d'Astronomia i Astrof\'{\i}sica, Universitat de
Val\`encia, C/ Dr. Moliner, 50, 46100, Burjassot, Valencian Country, Spain.\\ 
$^{2}$Observatori Astron\`omic, Universitat de Val\`encia, C/
Catedr\`atic Jos\'e Beltr\'an 2, 46980, Paterna, Valencian Country, Spain. 
}

\date{Accepted XXX. Received YYY; in original form ZZZ}

\pubyear{2015}

\begin{document}
\label{firstpage}
\pagerange{\pageref{firstpage}--\pageref{lastpage}}
\maketitle

\begin{abstract}
     Since Fanaroff \& Riley (1974) reported the morphological and brightness dichotomy of radiogalaxies, and it became clear that the symmetric emission from jets and counter-jets in the centre-brightened, less powerful, FRI sources could be caused by jet deceleration, many works have addressed different mechanisms that could cause this difference. Recent observational results seem to indicate that the deceleration must be caused by the development of small-scale instabilities that force mixing at the jet boundary. According to these results, the mixing layer expands and propagates down to the jet axis along several kiloparsecs, until it covers the whole jet cross-section. Several candidate mechanisms have been proposed as the initial trigger for the generation of such mixing layer. However, the instabilities proposed so far do not fully manage to explain the observations of FRI jets and/or require a triggering mechanism. Therefore, there is not still a satisfactory explanation for the original cause of jet deceleration. In this letter, I show that the penetration (and exit) of stars from jets could give the adequate explanation by means of creating a jet-interstellar medium mixing layer that expands across the jet.   
\end{abstract}




\section{Introduction}

    Extragalactic jets originate in the surroundings of supermassive black holes (SMBH) at the nuclei of active galaxies, and are probably generated by the extraction of rotational energy from the SMBH \citep{bz77}. These astrophysical objects manage to propagate from those compact regions to scales that are orders of magnitude larger than the formation region. We know now that, depending mainly on their power, but also on environmental conditions, jets can have different large-scale morphologies. \citet{fr74} established the well known brightness and morphological dichotomy of radiogalaxies, with brighter and edge-brightened type IIs (FRII) corresponding to jets showing hotspots and lobes at large scales, and type Is (FRI) corresponding to those showing weaker emission and irregular, plumed structures. Later, \citet{bi84,bi86a,bi86b} developed the paradigm of jet deceleration in FRIs: a mixing layer developing from the jet-ambient medium contact surface and mass-loading the flow down to the jet axis would favour deceleration to transonic, sub-relativistic velocities, and decollimation. 
    
        In \citet{lb14} (see also references therein), the authors summarise the properties of a number of FRI jets derived from detailed observations and modelling, and concluded that the deceleration process is strongly dissipative and that it develops from the jet boundary to its axis. This result favours small-scale instabilities as the cause of deceleration. 
            
    Among the proposed mechanisms to explain the generation of a mixing layer, we find the development of Kelvin-Helmholtz (KH) or current-driven (CD) instability modes \citep{pe05,pe10,ro08,tb16,ma16,mi17,ma19}, or strong recollimation shocks \citep{pm07}. However, neither signatures of long-wavelength, disruptive instability modes, nor of strong recollimation shocks are observed in FRI jets \citep[e.g.,][]{lb14}.\footnote{See, however, the development of a large-scale, helical structure well inside the radio jet of NGC 315 \citep{wo07} -although the nature of this structure is unclear-, and the case of M87, where the jet is decollimated beyond knot A, probably disrupted by the growth of KH instability modes to nonlinear amplitudes \citep[e.g.,][]{he11}} Small-scale instabilities that produce mixing layers can be associated to Kelvin-Helmholtz instabilities in rotating flows \citep{mk07,me08,mk09}, Rayleigh-Taylor \citep{mm13,ma17,to17}, or centrifugal instabilities \citep{gk18a,gk18b} developing in expanding and recollimating jets. Despite the lack of observational signatures for strong recollimation shocks, such instabilities could develop in expanding and smoothly recollimating outer jet layers, which is reported in \citet{lb14}. Still, a perturbing mechanism would be required to trigger their growth. Moreover, the smooth recollimation of the outer jet surface takes place after the entrainment has already started \citep[see Fig.~25 in][]{lb14}.
    
      Mass-load by stellar winds was also proposed as a decelerating mechanism by \citet{ko94}, who showed that the interaction between jets and stars could be treated as a hydrodynamical problem and anticipated that low power FRI jets ($L_{\rm j}\leq 10^{42}\,{\rm erg\,s^{-1}}$) could be efficiently decelerated by a stellar population with winds of $\dot{M}_{\rm w} \sim 10^{-12}\,{\rm M_\odot\,yr^{-1}}$. Later, \citet{bo96} confirmed this result by means of steady-state jet solutions. This result was also tested by \citet{pe14} using dynamical simulations. \citet{hb06} also showed that powerful enough stellar winds could decelerate low power jets. \citet{pe14} pointed out that such mass-load rates would be an inefficient mechanism for jet deceleration in the case of archetypical FRI jet powers, $L_{\rm j}\simeq 10^{43}-10^{44}\,{\rm erg\,s^{-1}}$. Furthermore, the mass-load by stellar winds would translate into a homogeneous decelerating process, in contradiction with the reported boundary-to-axis deceleration in \citet{lb14} for these powerful FRIs.
                  
     Altogether, the origin of the mixing layer that drives the deceleration mechanism of large-scale FRI jets is unknown, although it seems clear that the triggering mechanism for the formation of this mixing, turbulent layer has to be necessarily caused by small-scale perturbations that contribute to jet deceleration via dissipative processes, such as shocks and turbulent mass-load from the jet boundary. In this letter, I propose, the penetration (exit) of stars in (from) jets as a plausible scenario to trigger jet mixing with the interstellar medium (ISM) and thus explain the observed properties of decelerating jets, based on a collection of results already published and basic calculations. 
     
     The letter is organized as follows: In Section~2, the mechanism that triggers mixing is presented, whereas Section~3 includes a discussion about this idea and its implications. Finally, in Section~4 I present the conclusions of this work.

\section{Star penetration and exit as the origin of mixing}
   
\subsection{Mixing scales in FRI jets}   
  In the previous section I have reviewed a collection of results that indicate the presence of a mixing layer surrounding FRI jets in the inner kiloparsecs. This layer expands from the boundary to the axis, from an axial distance of $\sim 100\,$~pc to several kiloparsecs from the active nucleus, as modelled from VLA observations \citep{lb02,wa09,lb14}. The question is then what is the mechanism that triggers the formation of such a mixing region? Whatever this mechanism is, it has to be caused by a small-scale (because it has been impossible to observe it so far) and basically continuous (in order to produce the observed large-scale structures) process. Taking into account that different types of instabilities have been proposed, but they do not seem to fit the observations, I study here the possibility that stars penetrating and exiting the jet from the jet base to $\sim 0.1 - 1$~kpc can continuously perturb the jet boundary and generate a mixing layer. Actually,  \citet{pe17} have shown, via three-dimensional numerical simulations, that the contact of a star-wind system with the jet boundary produces a shock wave that propagates upstream along the jet shear-layer, and a conical shock that propagates downstream, through the jet, with the expected Mach angle. Downstream of the position of the star, a turbulent mixing region is developed, which is formed, not only by shocked stellar wind, but also by ambient gas, and widens the initial jet shear-layer \citep[see Figs. 3 and 5-8 in][]{pe17}. Strong dissipation is expected from this process, leading to limb-brightening of the jet at the interaction locations. 
  
  Star interaction with the jet boundaries can be regarded as a valid triggering mechanism as long as the time-scale of star penetration or the time between interactions is short enough compared to the dynamical time-scale of development of the mixing layer, which is basically determined by the time required for the layer to reach the jet axis. 
  
     Although there are no estimates of the velocities of spreading of mixing layers in relativistic flows, \citet{dy93} suggested that, as in the case of classical flows, the opening angle of the boundary layer is $\propto \sqrt{(\rho_{\rm ISM}/\rho_{\rm j})} M^{-1}$, where $M$ is the jet Mach number. \citet{ko94} also suggested that the turbulent layer formed between the cometary tail of shocked stellar winds and shocked jet flow in jet-star interactions would initially develop as ${\cal M}^{-1}$ (relativistic Mach number).\footnote{Note, however, that I suggest in this letter that stars act, not only as a source of mass, but also, and principally, as a source of perturbations at the jet boundary that enhance mixing between the jet and the ISM gas.} The inverse dependence of layer expansion with the Mach number implies that the mixing region will expand faster (with larger opening angles) into jets that are not dominated by the kinetic energy (rest-mass density and velocity), as expected for relatively low power FRI jets. In addition, \citet{pe05,pe10} studied the development of KH instability modes with various wavelengths in jets with different properties, via two and three-dimensional simulations. The authors found that the spreading rates of the mixing layer during the post-linear growth of the modes follow the ${\cal M}^{-1}$ (relativistic Mach number) trend, and the expansion velocities are in all cases $\sim 10^{-4}-10^{-2}\,c$. However, the jet-to-ambient density ratio in those jets is probably large ($\rho_{\rm j}/\rho_{\rm a} = 0.1$) to be compared with FRIs, in which this ratio is expected to be much smaller \citep{lb02,pm07}. Finally, according to the modelling of FRI jet deceleration by \citet{wa09} \citep[based on][for the jet in 3C~31]{lb02}, the layer crosses $\sim 100\,{\rm pc}$ along an axial distance of $\sim 3\,{\rm kpc}$, which gives an angle with the jet axis of $\sim 2^\circ$, and a ratio of the spreading rate to the jet velocity of $0.03\,c$. In this case, the time-scale of complete mixing would be $\sim 10^4$~yr. 
     
     In conclusion, the mixing scales are $\sim 10^4$~yr and several kiloparsecs. In the next subsection, I show that both the temporal and the spatial scales of star penetration and exit are much shorter than those global deceleration scales.

  \subsection{Interaction rates and scales} \label{sec:ias}
  
  The number of stars entering (exiting) the jet per unit time and distance, ${\cal N}_{\rm p}$, can be approximated as
   
   \begin{equation}
   {\cal N}_{\rm p}\, \sim \, n_{\rm s}(z)\,v_{\rm s}(z)\,\frac{S_{\rm j}(z)}{z}, 
   \end{equation}
where $z$ is the axial cylindrical coordinate, $S_{\rm j}(z) \sim \pi R_{\rm j}(z)\,z$ is half of the conical jet surface (the one opposing the direction of stellar rotation around the active nucleus, with $R_{\rm j}(z)$ the jet radius), and $n_{\rm s}(z)$ and $v_{\rm s}(z)$ are the stellar number density and rotation velocity in the considered region, respectively. At parsec scales, we obtain: 
       \begin{equation}
   {\cal N}_{\rm p}\, \sim \, 3 \times 10^{-4} \left(\frac{n_{\rm s}(z)}{1\, {\rm pc^{-3}}}\right) \,\left(\frac{v_{\rm s}(z)}{10^7\, {\rm cm\,s^{-1}}}\right)\,\left(\frac{R_{\rm j}(z)}{1\,{\rm pc}}\right) \,{\rm pc^{-1}\,yr^{-1}}. 
   \end{equation}
          
    This means that if the stellar density is of the order of $0.1-1\,{\rm pc^{-3}}$ \citep[see, e.g.,][]{ara13,vi17,tabr19}, and the jet radius is $\sim 10\,{\rm pc}$ , there is one star penetrating (and exiting) the jet every $\sim 3\times 10^{2}-10^{3}\,{\rm yr}$ per jet radius unit distance. 
   
    There is a relevant correction that needs to be made to the estimate given above: stars do not penetrate (exit) the jets instantaneously. The penetration time, $t_p$, is given, in the absence of a shear layer, by the size of the shocked stellar wind bubble that surrounds the star in the interstellar medium \citep[ISM, $10^{14}-10^{16}\,{\rm cm}$, depending on the type of star, see, e.g.,][]{pe17,tabr19} and the interaction region at the exit, once the bubble has been eroded as the star crosses the jet \citep[see][]{br12}. The size of the interaction region at equilibrium, $R_{\rm int}$, is given by the position at which the stellar wind and jet linear momenta become equal \citep[as measured from the stellar centre,][]{ko94}:
 \begin{eqnarray} \label{eq:rint}
  R_{\rm int} = 2.14 \times 10^{12} 	\left(\frac{\dot{M}_{\rm w}}{10^{-11} {\rm M_\odot\,yr^{-1}}}\right)^{1/2} \left(\frac{v_{\rm w}}{10 {\rm km\,s^{-1}}}\right)^{1/2} \times \nonumber \\ \left(\frac{L_{\rm j}}{10^{43} {\rm erg\, s^{-1}}}\right)^{-1/2} \left(\frac{v_{\rm j}}{c}\right)^{-1/2} \left(\frac{h_{\rm j}}{c^2}\right)^{1/2} \left(\frac{R_{\rm j}}{1 \,{\rm pc}}\right)\,{\rm cm}, 
\end{eqnarray}             
where $\dot{M}_{\rm w}$ and $v_{\rm w}$ are the wind mass-loss and velocity, $L_{\rm j}$ is the jet total power, $v_{\rm j}$ is the jet velocity, and $h_{\rm j}$ is its enthalpy per unit mass. Therefore, $t_{\rm p}\sim 1-10^3\,{\rm yr}$, for the given typical rotation velocity, $v_{\rm s}\sim 10^{7}\,{\rm cm\,s^{-1}}$ (and $t_{\rm exit}\leq 1\,{\rm yr}$). 

   If we also take into account the presence of a shear-layer surrounding the jet with a thickness of 1-10\% of the jet radius (with $R_{\rm j}\sim 1-10$~pc), as deduced from the transversal structure inferred from observations and modelling \citep[see, e.g.][for a recent study on the quasar S5~0836+710 and Schulz et al., submitted, for the case of 3C~111]{vg19}, the penetration (and exit) time of a single star\footnote{The penetration and exit times would be even longer for binary or multiple systems.} becomes of the order of $10^{17-18}/10^{7}\, {\rm s}\sim 10^{3-4}\,{\rm yr}$, i.e., longer than the penetration/exit rate ($\sim 3\times 10^{2}-10^{3}\,{\rm yr}$ for $R_{\rm j}\sim 10\,{\rm pc}$, see above).
      
    In conclusion, the penetration and exit of stars in the jet might be a continuous phenomenon in time, with one star continuously entering (exiting) the jet every 1-10~pc. Therefore, the triggering of jet-ISM mixing by stars can be considered as a continuous process taking place in the inner jet regions and developing along the typical, orders of magnitude larger deceleration scales given by \citet{lb14}, i.e., 3-10~kpc.
        
      \section{Discussion}    
      
 \subsection{The development of the mixing layer}
 
    In this section I discuss how the jet-ISM mixing layer could be formed from the continuous crossing of the jet boundaries by obstacles. The impact of stars at different azimuthal locations and the toroidal velocity of the jet shear-flow would favour the spreading of the layer around the jet perimeter. In this respect, it is reasonable to expect that the toroidal velocity of expansion of the mixing layer is of the order of the radial and axial velocities, i.e., $0.03\,c$ (see previous Section). Thus, the time needed by the mixing region to spread around half of the jet surface is $\simeq 30\,\pi\,R_{\rm j}$, which gives $\sim 10^2-10^3$~yr for $R_{\rm j}\,=\,1-10$~pc (the expected jet radius at $z\sim 10-100\,{\rm pc}$ for an opening angle of 0.1~rad). This time (computed for a single interaction) is also shorter than the deceleration time-scale. In addition, the interactions taking place at different locations along the jet (i.e., in the jet flow direction) would enhance turbulent mixing and deceleration at progressively closer distances to the axis, as the mixing layer expands.

    Moreover, the computed value of ${\cal N}_{\rm p}$ represents a lower limit of obstacle-jet interaction rate per unit length, because we have not taken into account clouds in the narrow-line region, which can enter the jet if dense or large enough. In general, an obstacle will be able to enter the jet if the orbital velocity is larger than the sound speed in the shocked obstacle \citep[see, e.g.,][]{br12}, i.e., $v_{\rm o} \geq v_{\rm sc}$, with $v_{\rm sc} \sim c \sqrt{\rho_{\rm j}\,\gamma_{\rm j}/\rho_{\rm o}}$, where $c$ is the speed of light, $\gamma_j=1/\sqrt{1-(v_{\rm j}/c)^2}$ is the jet Lorentz factor, and $\rho_{\rm j}$ and $\rho_{\rm o}$ are the jet and obstacle densities in the reference frame of the obstacle. This implies that dense enough clouds within the narrow line region (i.e., up to 1~kpc) can penetrate the jet, crossing the slowed down mixing layer, once it has started to decelerate. Taking $v_{\rm o}\simeq 10^7\, {\rm cm\,s^{-1}}$,  this condition requires $\rho_{\rm j}/\rho_{\rm o} < 10^{-7}$, for small Lorentz factors. Nevertheless, independently of whether the clouds penetrate or not into the jet, they also certainly represent a perturbation of the jet boundary.

       In summary, stars and possibly clouds could trigger the formation of a mixing layer and jet deceleration. In addition, mass-load by stellar winds can contribute to internal deceleration as the stars cross the jet, depending on the jet power and properties \citep[][, Perucho et al., in preparation]{pe14}. Actually, progressive jet deceleration and dissipation can explain that the surface separating the jet and the mixing layer has a parabolic shape (peaked at the point where it reaches the jet axis) in the modelling performed by \citet{wa09}, and also for some sources in \citet{lb14}, on the basis of the steepening of the Mach angle due to both the decrease in jet velocity and the increase in sound speed. The stellar winds could provide such an internal, mild deceleration mechanism.

\subsection{Possible coupling to small-scale instability modes}     
   Although the interactions are locally non-linear \citep{pe17}, if we consider each of them as a linear perturbation to the whole jet structure, they could couple to short-wavelength instability modes while advected downstream, and be maintained by subsequent interactions. In other words, the instability would be fed by new stars entering and exiting the jet.  

     The expected wavelength generated by such an interaction is of the order of the size of the obstacle, i.e., $\ll 1$~pc, for the sizes of the stellar bubbles and interaction regions given above (from $\leq 0.1\,{\rm pc}$ at penetration to $\leq 10^{-4}\,{\rm pc}$ for high-mass stars or red giants at the exit). Note, however, that the interaction with binary stars would produce larger interaction regions. The different obstacles would thus produce a spectrum of small-scale perturbation wavelengths of the boundary. 
     
     Altogether, these perturbations could trigger short shear (Kelvin-Helmholtz) instability modes, which typically show high growth-rates \citep[see, e.g.][]{ha00,pe05,pe07}, or Rayleigh-Taylor/centrifugal instability as the jets expand \citep{mm13,ma17,to17,gk18a,gk18b}. Nevertheless, the continuous process of star penetration and exit could, by itself, generate the formation of a turbulent mixing layer without needing to couple to instability modes. This scenario should be tested by numerical simulations, but one has to keep in mind that the difference between the perturbation and the deceleration scales is too large to be affordable in computational terms. As a consequence, these tests remain out of the scope of this work and should be left for the future.
          
\subsection{Observability}      

     The interactions described here should generate irregularities in the jet boundary (knots and filaments), too small in general to be directly observed: the value of $R_{\rm int}$ derived above implies an angular size of $\sim 10^{-2}\,\mu$as in the nearby radiogalaxy M87, whereas the largest bubble sizes could represent  $\sim 0.1~$mas. This latter size is slightly below resolution at 15-43~GHz, \citep{ko07,wa18}, and about the one achieved at 86~GHz \citep{ki18}. However, the jet region observed at 86~GHz is $\leq 1$~pc, making the interaction with a massive star less probable at a given time. Therefore, an increase of the resolution (to resolve the interaction region) and sensitivity (to increase the observed jet length and thus increase the probability of detection) would be needed in order to observe one of the expected imprints of these interactions in nearby sources. Still, the limb brightening and onset of a shear layer at parsec scales \citep[see, e.g.,][]{me16} could well be related to interactions with stars and/or clouds \citep[as it was suggested for the knots in this same source by][]{bk79}.  
     
     As a consequence, only large-scale interaction regions, as those produced by stars with high mass-loss rates, binary systems and/or shocked-wind bubbles at the jet surface and preferentially far from the nucleus (where the jet velocity is smaller and the radius is larger, see Eq.~\ref{eq:rint}) could be observed as knots or bow structures, as it has been suggested for the jet in Centaurus~A \citep[see, e.g.,][]{wo08,go10,wy13,mu14}. Only the large-scale effect, namely a continuous, growing boundary layer could be observed.

 \subsection{Brightness flaring}      
         
      According to \citet{lb14}, there is an apparently discontinuous brightness flaring of FRI jets at different wavelength ranges, before the main process of jet deceleration starts. This happens at $\sim 1\,{\rm kpc}$ from the galactic nucleus, and approximately where the jet is expanding with the largest opening angle (geometrical flaring), coinciding with the drop in ambient pressure. Their model results in half of the sources showing an increase in brightness possibly as a consequence of rapid jet expansion, as long as the emissivity does not fall too fast, according to the authors. Within the idea that this letter presents, this could be explained by the increase of the number of stars in the jet when the fast expansion takes place: The total dissipated and radiated power is proportional to the number of interactions taking place at a given time, $N_{\rm int}\propto n_{\rm s}(z) \,V_{\rm j} \propto n_{\rm s}(z) \,R_{\rm j}(z)^2$. In the flaring region, the jet radius grows with distance to the source with an exponent, $\kappa>1$ \citep[in][the authors use a third order polynomial to fit the evolution of the jet radius in this region]{lb14} so the dissipated energy will grow with distance while the drop in $n_{\rm s}(z)$ is shallower than $1/z^{2\,\kappa}$. The sudden jet expansion can be caused by a combination of a drop in the ambient pressure (as reported in that paper) and an increase in jet pressure, also caused by dissipation \citep[e.g.,][Perucho et al., in preparation]{bo96,pe14}. It is therefore reasonable to expect a maximum in emissivity related to the dissipation produced by jet-star interactions, which would result in the observed jet brightening at distances $\sim 1\,{\rm kpc}$ \citep[see][for works on the collective role of jet star interactions in jet brightness at high energies]{wy15,vi17}\footnote{See also \citet{rl18,pe19} for recent reviews that include discussions on the radiative output of jet-star interactions (and references therein)}. Farther downstream, the number of stars may fall fast with distance to the galactic nucleus and brightness be maintained by the enhanced turbulence and small-scale shocks within the mixing layer and the jet cross-section. Therefore, jet-star interactions can explain the brightness flaring of FRI jets, by simple accumulation of embedded stars in a region of fast jet expansion.

\subsection{The FRII case}
      In the case of FRII jets, which are typically more powerful than FRIs \citep[see, e.g.,][]{gc01}, the role of stars in the formation a developing a mixing layer is necessarily weaker because of three main reasons: 1) there is basically no internal deceleration by stellar winds \citep[see, e.g.,][]{pe14}, so the development of a possible mixing layer towards the jet axis would be much slower, 2) small-scale instability modes in powerful jets could lead to the growth of resonant KH modes that stabilize the jet spine \citep{pe05,pe07}, and 3) $v_{\rm sc}$ (see Sect.~\ref{sec:ias}) is much larger in this case, which difficults that $v_{\rm o} \geq v_{\rm sc}$), so that smaller quantities of mass-load will be deposited inside the jet, and the impact of obstacles will be minimized by erosion as they penetrate the jet.

\section{Conclusions}
     Summarising, the penetration and exit of stars in and out from jets is a process that can account for the observed properties of jet deceleration in FRI radiogalaxies, by triggering the formation of a turbulent jet-ISM mixing layer at the jet surface that is continuously fed by ongoing interactions. Therefore, the scenario proposed here could solve the problem of the lack of observational evidence for triggering mechanisms that result in the development of the mixing layer that is demanded by the FRI evolution paradigm \citep{bi84} and models derived from observations \citep{lb14}. This solution is compatible with the lack of strong recollimation shocks, edge-brightening at the inner regions, geometrical and brightness flaring, and boundary-to-axis deceleration, as reported by \citet{lb14}.  
  
         A quantitative approach to the idea presented here obviously requires numerical simulations, owing to the non-linear nature of the interactions and the generation of the mixing layer. However, the different spatial scales of the jet-star interaction as compared to the jet deceleration scales will difficult the set up such simulations. Despite this difficulty, future work should be aimed to reproduce and test the proposed scenario.

\section*{Acknowledgements}
This work has been supported by the Spanish Ministerio de Econom\'{\i}a y Competitividad
(grant AYA2016-77237-C3-3-P) and the Generalitat Valenciana (grant PROMETEU/2019/071). I thank the anonymous referee of this work for his/her positive comments and suggestions.


\begin{thebibliography}{99}

\bibitem[\protect\citeauthoryear{Araudo, Bosch-Ramon \& Romero}{2013}]{ara13} Araudo, A.T., Bosch-Ramon, V., Romero, G. 2013, MNRAS, 436, 3626
\bibitem[\protect\citeauthoryear{Bicknell}{1984}]{bi84} Bicknell G.V., 1984, ApJ, 286, 68
\bibitem[\protect\citeauthoryear{Bicknell}{1986a}]{bi86a} Bicknell G.V., 1986a, ApJ, 300, 591
\bibitem[\protect\citeauthoryear{Bicknell}{1986b}]{bi86b} Bicknell G.V., 1986b, ApJ, 305, 109

\bibitem[Blandford \& Znajek(1977)]{bz77} Blandford, R.D, Znajek, R. 1977, MNRAS, 179, 433
\bibitem[Blandford \& Koenigl(1979)]{bk79} Blandford, R.D, Koenigl, A. 1979, ApL, 20, 15
\bibitem[\protect\citeauthoryear{Bosch-Ramon, Perucho \& Barkov}{2012}]{br12} Bosch-Ramon, V., Perucho, M., Barkov, M.V., 2012, A\&A, 539A, 69 
\bibitem[\protect\citeauthoryear{Bowman, Leahy \& Komissarov}{Bowman et al.}{1996}]{bo96} Bowman M., Leahy J.P., Komissarov S.S., 1996, MNRAS, 279, 899
\bibitem[\protect\citeauthoryear{Celotti \& Ghisellini}{2008}]{cg08} Celotti A., Ghisellini G., 2008, MNRAS, 385, 283
\bibitem[\protect\citeauthoryear{De Young}{1993}]{dy93} De Young D.S., 1993, ApJ, 405, L13

\bibitem[\protect\citeauthoryear{Fanaroff \& Riley}{1974}]{fr74} Fanaroff B.L., Riley J.M., 1974, MNRAS, 167, 31
\bibitem[\protect\citeauthoryear{Ghisellini \& Celotti}{2001}]{gc01}  Ghisellini, G., Celotti, A. 2001, A\&A, 379, L1

\bibitem[\protect\citeauthoryear{Goodger et al.}{2010}]{go10} Goodger, J.L., Hardcastle, M., Croston, J.H., et al. 2010, ApJ, 708, 675
\bibitem[\protect\citeauthoryear{Gourgouliatos \& Komissarov}{2018a}]{gk18a} Gourgouliatos K.N., Komissarov, S.S., 2018a, Nature Astronomy, 2, 167
\bibitem[\protect\citeauthoryear{Gourgouliatos \& Komissarov}{2018b}]{gk18b} Gourgouliatos K.N., Komissarov, S.S., 2018b, MNRAS, 475, 125
\bibitem[\protect\citeauthoryear{Hardee}{2000}]{ha00} Hardee P.E., 2000, ApJ, 533, 176
\bibitem[\protect\citeauthoryear{Hardee \& Eilek}{2011}]{he11} Hardee P.E., Eilek, J.A., 2011, ApJ, 735, 61
\bibitem[\protect\citeauthoryear{Hubbard \& Blackman}{2006}]{hb06} Hubbard A., Blackman E.G., 2006, MNRAS, 371, 1717
\bibitem[\protect\citeauthoryear{Kim et al.}{2018}]{ki18} Kim, J.-Y., Krichbaum, T.P., Lu, R.-S., et al., 2018, A\&A, 616A, 188

\bibitem[\protect\citeauthoryear{Komissarov}{1994}]{ko94} Komissarov S.S., 1994, MNRAS, 269, 394
5
\bibitem[\protect\citeauthoryear{Kovalev et al.}{2007}]{ko07} Kovalev, Y.Y., Lister, M.L., Homan, D.C., Kellermann, K.I., 2007, ApJL, 668, 27 
\bibitem[\protect\citeauthoryear{Laing \& Bridle}{2002}]{lb02} Laing R.A., Bridle A.H., 2014, MNRAS, 336, 1161
\bibitem[\protect\citeauthoryear{Laing \& Bridle}{2014}]{lb14} Laing R.A., Bridle A.H., 2014, MNRAS, 437, 3405
\bibitem[\protect\citeauthoryear{Massaglia et al.}{2016}]{ma16} Massaglia, S., Bodo, G., Rossi, P., Capetti, S., Mignone, A. 2016, A\&A, 596A, 12
\bibitem[\protect\citeauthoryear{Massaglia et al.}{2019}]{ma19} Massaglia, S., Bodo, G., Rossi, P., Capetti, S., Mignone, A. 2019, A\&A, 621A, 132
\bibitem[\protect\citeauthoryear{Matsumoto \& Masada}{2013}]{mm13} Matsumoto, J., Masada, Y. 2013, ApJ, 772, L1.
\bibitem[\protect\citeauthoryear{Matsumoto et al.}{2017}]{ma17} Matsumoto, J., Aloy, M.A., Perucho, M. 2017, ApJ, 472, 1421.
\bibitem[\protect\citeauthoryear{Meliani \& Keppens}{2007}]{mk07} Meliani, Z., Keppens, R., 2007, A\&A, 475, 785.
\bibitem[\protect\citeauthoryear{Meliani, Keppens \& Giacomazzo}{Meliani et al.}{2008}]{me08} Meliani Z., Keppens R., Giacomazzo B. 2008, A\&A, 491, 321.
\bibitem[\protect\citeauthoryear{Meliani \& Keppens}{2009}]{mk09} Meliani, Z., Keppens, R. 2009, ApJ, 705, 1594.
\bibitem[\protect\citeauthoryear{Mertens et al.}{2016}]{me16} Mertens, F., Lobanov, A.P., Walker, R.C., Hardee, P. E. 2016, A\&A, 595A, 54M
\bibitem[\protect\citeauthoryear{Millas, Keppens \& Meliani}{2017}]{mi17} Millas, D., Keppens, R., Meliani, Z. 2017, MNRAS, 470, 592
\bibitem[\protect\citeauthoryear{M\"uller et al.}{2014}]{mu14} M\"uller, C., Kadler, M., Ojha, R., et al., 2014, A\&A, 569A, 115
\bibitem[\protect\citeauthoryear{Perucho, Mart\'{\i} \& Hanasz}{Perucho et al.}{2005}]{pe05} Perucho M.. Mart\'{\i} J.M., Hanasz M., 2005, A\&A, 443, 863
\bibitem[\protect\citeauthoryear{Perucho \& Mart\'{\i}}{2007}]{pm07} Perucho M., Mart\'{\i} J.M., 2007, MNRAS, 382, 526
\bibitem[\protect\citeauthoryear{Perucho et al.}{2007}]{pe07} Perucho, M., Hanasz, M., Mart\'{\i}, J.M., Miralles, J.A. 2007, Phys. Rev. E, 75, 056312
\bibitem[\protect\citeauthoryear{Perucho et al.}{2010}]{pe10} Perucho M., Mart\'{\i} J.M., Cela, J.M., Hanasz, M., de la Cruz, R. Rubio, F., 2010, A\&A, 519, A41
\bibitem[\protect\citeauthoryear{Perucho, Mart\'{\i}, Laing, Hardee}{Perucho et al.}{2014}]{pe14} Perucho M., Mart\'{\i} J.M., Laing R.A., Hardee P.E. 2014, MNRAS, 441, 1488
\bibitem[\protect\citeauthoryear{Perucho, Bosch-Ramon \& Barkov}{2017}]{pe17} Perucho, M., Bosch-Ramon, V., Barkov, M.V., 2017, A\&A, 606A, 40 
\bibitem[Perucho(2019)]{pe19} Perucho, M., 2019, Galaxies, 7, 70
\bibitem[Rieger \& Levinson(2018)]{rl18} Rieger, F.M., Levinson, A. 2018, Galaxies, 6, 116
\bibitem[\protect\citeauthoryear{Rossi et al.}{2008}]{ro08} Rossi P., Mignone A., Bodo G., Massaglia S., Ferrari A., 2008, A\&A, 488, 795
\bibitem[\protect\citeauthoryear{Tchekhovskoy \& Bromberg}{2016}]{tb16} Tchekhovskoy A., Bromberg O., 2016, MNRAS, 461, L46
\bibitem[\protect\citeauthoryear{Toma, Komissarov \& Porth}{Toma et al.}{2017}]{to17} Toma, K., Komissarov, S.S., Porth, O. 2017, MNRAS, 472, 1253.
\bibitem[\protect\citeauthoryear{Torres-Alb\`a \& Bosch-Ramon}{2019}]{tabr19} Torres-Alb\`a, N., Bosch-Ramon, V. 2019, A\&A, 623, A91
\bibitem[\protect\citeauthoryear{Vega-Garc\'{\i}a, Perucho \& Lobanov}{2019}]{vg19} Vega-Garc\'{\i}a, L., Perucho, M., Lobanov, A. 2019, A\&A, 627, A79
\bibitem[\protect\citeauthoryear{Vieyro, Torres-Alb\`a \& Bosch-Ramon}{2017}]{vi17} Vieyro, F.L., Torres-Alb\`a, N., Bosch-Ramon, V., 2017, A\&A, 604A, 57\bibitem[\protect\citeauthoryear{Walker et al.}{2018}]{wa18} Walker, R.C., Hardee, P.E., Davies, F.B., Ly, C., Junor, W., 2018, ApJ, 855, 128 
\bibitem[\protect\citeauthoryear{Wang et al.}{2009}]{wa09} Wang Y., Kaiser C.R., Laing R., Alexander P., Pavlovski G., Knigge C., 2009, MNRAS, 397, 1113 
\bibitem[\protect\citeauthoryear{Worrall et al.}{2007}]{wo07} Worrall, D.M., Birkinshaw, M., Laing, R.A., Cotton, W.D., Bridle, A.H., 2007, MNRAS, 380, 2
\bibitem[\protect\citeauthoryear{Worrall et al.}{2008}]{wo08} Worrall, D.M., Birkinshaw, M., Kraft, R.P., et al., 2008, ApJ, 673, L135
\bibitem[\protect\citeauthoryear{Wykes et al.}{2013}]{wy13} Wykes S., Croston J.H., Hardcastle M.J. et al., 2013, A\&A, 558A, 19
\bibitem[\protect\citeauthoryear{Wykes et al.}{2015}]{wy15} Wykes S., Hardcastle M.J., Karakas A.I., Vink J.S., 2015, MNRAS, 447, 1001



\end{thebibliography}





\bsp	
\label{lastpage}
\end{document}